\documentclass[11pt]{amsart}

\usepackage{amssymb,amsthm,amsmath,amssymb,xcolor}
\usepackage{enumerate}
\usepackage[colorlinks]{hyperref}

\newcommand{\dd}{\mathrm{d}}
\newcommand{\E}{\mathbb{E}}
\newcommand{\1}{\textbf{1}}
\newcommand{\R}{\mathbb{R}}
\newcommand{\C}{\mathbb{C}}

\newcommand{\Ent}[1]{h\left( #1 \right)}

\newcommand{\po}[2]{\frac{\textrm{d} #1}{\textrm{d} #2}}
\newcommand{\e}{\varepsilon}

\usepackage[paper=a4paper, left=1.3in, right=1.3in, top=1.2in, bottom=1.2in]{geometry}
\newtheorem{theorem}{Theorem}

\theoremstyle{remark}

\theoremstyle{definition}

\begin{document}

\title{Two remarks on generalized entropy power inequalities}
\author{Mokshay Madiman}
\address{University of Delaware}
\email{madiman@udel.edu}

\author{Piotr Nayar}
\address{University of Warsaw}
\email{nayar@mimuw.edu.pl}

\author{Tomasz Tkocz}
\address{Carnegie Mellon University}
\email{ttkocz@math.cmu.edu}

\thanks{M.M. was supported in part by the U.S. National Science Foundation through the grant DMS-1409504. 
P.\!~N. was partially supported by the National Science Centre Poland grant
2015/18/A/ST1/00553. The research leading to these results is part of a project that has received funding from the 
European Research Council (ERC) under the European Union's Horizon 2020 research and innovation programme (grant agreement No 637851).
This work was also supported by the NSF under Grant No. 1440140, while the authors were in residence at the Mathematical Sciences Research Institute 
in Berkeley, California, for the ``Geometric and Functional Analysis'' program during the fall semester of 2017.}

\begin{abstract}
This note contributes to the understanding of generalized entropy power inequalities.
Our main goal is to construct a counter-example regarding monotonicity and entropy comparison of weighted sums of independent 
identically distributed log-concave random variables. We also present a complex analogue of a recent dependent
entropy power inequality of Hao and Jog, and give a very simple proof.
\end{abstract}

\maketitle

{\footnotesize

}
\bigskip

\section{Introduction}

The differential entropy (or simply {\it entropy}, henceforth, since we have no need to deal with
discrete entropy in this note) of a random vector $X$ with density $f$ is defined as
\[
\Ent{X} = -\int_{\R^d} f \log f,
\]
provided that this integral exists. When the variance of a real-valued
random variable $X$ is kept fixed, it is a long known fact \cite{Bol1896} that
the differential entropy is maximized by taking $X$ to be Gaussian.
A related functional is the {\it entropy power} of $X$, defined by
$
N(X)=e^{\frac{2h(X)}{d}}.
$
As is usual, we abuse notation and write $h(X)$ and $N(X)$,
even though these are functionals depending only on the density
of $X$ and not on its random realization.

The entropy power inequality is a fundamental inequality in both Information Theory and Probability,
stated first by Shannon \cite{Sha48} and proved by Stam \cite{Sta59}. 
It states that for any two independent random vectors $X$ and $Y$ in $\R^d$ such that the entropies of $X, Y$ and $X+Y$ exist, 
$$
N(X+Y) \geq N(X) + N(Y) .
$$
In fact, it holds without even assuming the existence of entropies as long as we set an entropy power to 0 whenever the corresponding entropy does not exist,
as noted by \cite{BC15:1}.
One reason for the importance of this inequality in Probability Theory comes from its close connection 
to the Central Limit Theorem (see, e.g., \cite{Joh04:book, Mad17}). It is also closely related to
the Brunn-Minkowski inequality, and thereby to results in Convex Geometry and Geometric Functional Analysis (see, e.g., \cite{BM11:cras, MMX17:0}).

An immediate consequence of the above formulation of the entropy power inequality is its extension
to $n$ summands: if $X_1, \ldots, X_n$ are independent random vectors, then
$N(X_1+\ldots+X_n)\geq \sum_{i=1}^n N(X_i)$. 
Suppose the random vectors $X_i$ are not merely independent but also identically distributed,
and that $S_n=\frac{1}{\sqrt{n}} \sum_{i=1}^n X_i$; these are the normalized partial sums that 
appear in the vanilla version of the Central Limit Theorem. Then one concludes from the entropy power inequality 
together with the scaling property $N(aX)=a^2 N(X)$ that
$N(S_n)\geq N(S_1)$, or equivalently that 
\begin{equation}\label{eq:nto1}
h(S_n)\geq h(S_1).
\end{equation}

There are several refinements or generalizations of the inequality \eqref{eq:nto1} that one may consider.
In 2004, Artstein, Ball, Barthe and Naor \cite{ABBN04:1} proved (see \cite{MB06:isit, TV06, Shl07, Cou18:2} for simpler proofs 
and \cite{MB07, MG18} for extensions) that in fact, one has monotonicity of entropy along the  
Central Limit Theorem, i.e., $h(S_n)$ is a monotonically increasing sequence. If $N(0,1)$ is the standard normal distribution, Barron \cite{Bar86} had proved much earlier
that $h(S_n)\rightarrow h(N(0,1))$ as long as $X_1$ has mean 0, variance 1, and $h(X_1)>-\infty$.
Thus one has the monotone convergence of $h(S_n)$ to the Gaussian entropy, which is the maximum entropy possible under
the moment constraints. By standard arguments, the convergence of entropies is equivalent to the relative entropy between the distribution of $S_n$ and the
standard Gaussian distribution converging to 0,
and this in turn implies not just convergence in distribution but also convergence in total variation. This is the way in which entropy illuminates the Central Limit Theorem.

A different variant of the inequality  \eqref{eq:nto1}  was recently given by Hao and Jog \cite{HJ18},
whose paper may be consulted for motivation and proper discussion.
A random vector $X=(X_1,\ldots,X_n)$ in $\R^n$ is called {\it unconditional} if for every choice of 
signs $\eta_1,\ldots,\eta_n \in \{-1,+1\}$, the vector $(\eta_1X_1,\ldots,\eta_nX_n)$ 
has the same distribution as $X$. 
Hao and Jog \cite{HJ18} proved that if $X$ is an unconditional random vector in $\R^n$, then 
$\frac{1}{n}\Ent{X} \leq \Ent{\frac{X_1+\ldots+X_n}{\sqrt{n}}}$. If $X$ has independent and identically distributed
components instead of being unconditional, this is precisely  $h(S_n)\geq h(S_1)$ for real-valued random variables $X_i$ (i.e., in dimension $d=1$).

The goal of this note is to shed further light on both of these generalized entropy power inequalities. 
We now explain precisely how we do so.

To motivate our first result, we first recall the notion of Schur-concavity. One vector $a=(a_1,\ldots,a_n)$ in $[0,\infty)^n$ is 
{\it majorised} by another one $b=(b_1,\ldots,b_n)$, usually denoted $a\prec b$, if the nonincreasing rearrangements 
$a_1^*\geq\ldots\geq a_n^*$ and $b_1^*\geq\ldots\geq b_n^*$ of $a$ and $b$ satisfy the inequalities
$\sum_{j=1}^k a^*_j \leq \sum_{j=1}^k b_j^*$ for each $1\leq k\leq n-1$ and $\sum_{j=1}^n a_j = \sum_{j=1}^n b_j$. 
For instance, any vector $a$ with nonnegative coordinates adding up to $1$ is majorised by the vector $(1,0,\ldots,0)$ 
and majorises the vector $(\frac{1}{n},\frac{1}{n},\ldots,\frac{1}{n})$. 
Let $\Phi:\Delta_n\rightarrow \R$, where $\Delta_n=\{a\in [0,1]^n: a_1+\ldots+a_n=1\}$ is the standard simplex. 
We say that $\Phi$ is {\it Schur-concave} if $\Phi(a) \geq\Phi(b)$ when $a\prec b$.
Clearly, if $\Phi$ is Schur-concave, then one has $\Phi(\frac{1}{n},\frac{1}{n},\ldots,\frac{1}{n})\geq \Phi(a)\geq \Phi(1,0,\ldots,0)$
for any $a\in \Delta_n$.

Suppose $X_1,\ldots,X_n$ are i.i.d. copies of a random variable $X$ with finite entropy, and we define
\begin{equation}\label{eq:Phi}
\Phi(a)=\Ent{\sum \sqrt{a_i} X_i} 
\end{equation}
for $a\in\Delta_n$. Then the inequality  \eqref{eq:nto1} simply says that $\Phi(\frac{1}{n},\frac{1}{n},\ldots,\frac{1}{n})\geq \Phi(1,0,\ldots,0)$,
while the monotonicity of entropy in the Central Limit Theorem says that $\Phi(\frac{1}{n},\frac{1}{n},\ldots,\frac{1}{n})\geq \Phi(\frac{1}{n-1},\ldots,\frac{1}{n-1}, 0)$.
Both these properties would be implied by (but in themselves are strictly weaker than) Schur-concavity. Thus one is led to the natural question:
{\it Is the function $\Phi$ defined in \eqref{eq:Phi} a Schur-concave function?} For $n=2$, this would imply in particular
that $h(\sqrt{\lambda} X_1 +\sqrt{1-\lambda}X_2)$ is maximized over $\lambda\in [0,1]$ when $\lambda=\frac{1}{2}$.
The question on the Schur-concavity of $\Phi$ had been floating around for at least a decade, until  \cite{BNT16}
constructed a counterexample showing that $\Phi$ cannot be Schur-concave even for $n=2$.
It was conjectured in \cite{BNT16}, however, that for $n=2$, the Schur-concavity should hold if the random variable $X$
has a log-concave distribution, i.e., if $X_1$ and $X_2$ are independent, identically distributed, log-concave random variables, 
the function $\lambda \mapsto \Ent{\sqrt{\lambda}X_1+\sqrt{1-\lambda}X_2}$ should be nondecreasing on $[0,\frac{1}{2}]$. 
More generally, one may ask:
{\it if $X_1,\ldots,X_n$  are $n$ i.i.d. copies of a log-concave random variable $X$, is is true that
$\Ent{\sum a_iX_i} \geq \Ent{\sum b_iX_i}$ when $(a_1^2,\ldots,a_n^2) \prec (b_1^2,\ldots,b_n^2)$?}
Equivalently, is $\Phi$ Schur-concave when $X$ is log-concave?

Our first result implies that the answer to this question is negative.
The way we show this is the following:
since $(1,\frac{1}{n},\ldots,\frac{1}{n},\frac{1}{n}) \prec (1,\frac{1}{n-1},\ldots,\frac{1}{n-1},0)$, if Schur-concavity held, 
then the sequence $\Ent{X_1+\frac{X_2+\ldots+X_{n+1}}{\sqrt{n}}}$ would be nondecreasing and as it 
converges to $\Ent{X_1 + G}$, where $G$ is an independent Gaussian random variable with the same variance as $X_1$, 
we would have in particular that $\Ent{X_1+\frac{X_2+\ldots+X_{n+1}}{\sqrt{n}}} \leq \Ent{X_1 + G}$. We construct examples where the opposite holds.

\begin{theorem}\label{thm:schurfails}
There exists a symmetric log-concave random variable $X$ with variance $1$ such that 
if $X_0, X_1,\ldots$ are its independent copies and $n$ is large enough, we have
\[
\Ent{X_0 + \frac{X_1+\ldots+X_n}{\sqrt{n}}} > \Ent{X_0+Z},
\]
where $Z$ is a standard Gaussian random variable, independent of the $X_i$.
Consequently, even if $X$ is drawn from a symmetric, log-concave distribution, 
the function $\Phi$ defined in \eqref{eq:Phi} is not Schur-concave. 
\end{theorem}

Here by a {\it symmetric} distribution, we mean one whose density $f$ satisfies $f(-x)=f(x)$ for each $x\in\R$.

In contrast to Theorem \ref{thm:schurfails}, $\Phi$ does turn out to be Schur-concave if the distribution of $X$ is a symmetric Gaussian mixture, 
as recently shown in \cite{ENT18:1}. We suspect that Schur-concavity also holds for uniform distributions on intervals (cf. \cite{aim17:epi}).

Theorem \ref{thm:schurfails} can be compared with the afore-mentioned monotonicity of entropy property of the Central Limit Theorem.
It also provides an example of two independent symmetric log-concave random variables $X$ and $Y$ with the same variance 
such that $\Ent{X+Y} > \Ent{X+Z}$, where $Z$ is a Gaussian random variable with the same variance as $X$ and $Y$, 
independent of them, which is again in contrast to symmetric Gaussian mixtures (see \cite{ENT18:1}). 
The interesting question posed in \cite{ENT18:1} of whether, for two i.i.d. summands, 
swapping one for a Gaussian with the same variance increases entropy, remains open.

Our proof of Theorem \ref{thm:schurfails} is based on sophisticated and remarkable Edgeworth type expansions recently developed by
Bobkov, Chistyakov and G\"otze \cite{BCG13:rate}
en route to obtaining precise rates of convergence in the entropic central limit theorem, and is detailed in Section~\ref{sec:schurfails}.


The {\bf second contribution} of this note is an exploration of a technique to prove inequalities akin to the
entropy power inequality by using symmetries and invariance properties of entropy. It is folklore that when $X_1$ and $X_2$
are i.i.d. from a symmetric distribution, one can deduce the inequality $h(S_2)\geq h(S_1)$ in an extremely simple fashion
(in contrast to any full proof of the entropy power inequality, which tends to require relatively sophisticated machinery-- either going through
Fisher information or optimal transport or rearrangement theory or functional inequalities).
In Section~\ref{sec:proofcomplex}, we will recall this simple proof, and also deduce some variants of 
the inequality $h(S_2)\geq h(S_1)$ by playing with this basic idea of using invariance,
including a complex analogue of a recent entropy power inequality for dependent
random variables obtained by Hao and Jog \cite{HJ18}. 

\begin{theorem}\label{thm:complex}
Let $X = (X_1,\ldots,X_n)$ be a random vector in $\C^n$ which is complex-unconditional, 
that is for every complex numbers $z_1,\ldots,z_n$ such that $|z_j|=1$ for every~$j$, 
the vector $(z_1X_1,\ldots,z_nX_n)$ has the same distribution as $X$. Then
\[
 \frac{1}{n}\Ent{X} \leq \Ent{\frac{X_1+\ldots+X_n}{\sqrt{n}}}.
\]
\end{theorem}

Our proof of Theorem~\ref{thm:complex}, which is essentially trivial thanks to the existence of complex Hadamard matrices,
is in contrast to the proof given by \cite{HJ18} for the real case that proves a Fisher information inequality as an intermediary step.


We make some remarks on complementary results in the literature. Firstly, in contrast to the failure of Schur-concavity of $\Phi$
implied by Theorem \ref{thm:schurfails}, the function $\Xi:\Delta_n\rightarrow \R$ defined by
$\Xi(a)=\Ent{\sum a_i X_i}$  for i.i.d. copies $X_i$ of a random variable $X$, is actually Schur-convex when $X$ is log-concave \cite{Yu08:2}.
This is an instance of a reverse entropy power inequality, many more of which are discussed in \cite{MMX17:0}.
Note that the weighted sums that appear in the definition of $\Phi$ are relevant to the Central Limit Theorem because they have
fixed variance, unlike the weighted sums that appear in the definition of $\Xi$.

Secondly, motivated by the analogies with Convex Geometry mentioned earlier,
one may ask if the function $\Psi:\Delta_n\rightarrow \R$ defined by $\Psi(a)=\text{vol}_d(\sum_{i=1}^n a_i B)$,
is Schur-concave for any Borel set $B\subset\R^d$, where $\text{vol}_d$ denotes the Lebesgue measure on $\R^d$
and the notation for summation is overloaded as usual to also denote Minkowski summation of sets. 
(Note that unless $B$ is convex, $(a_1+a_2)B$
is a subset of, but generally not equal to, $a_1 B+a_2 B$.) The Brunn-Minkowski inequality
implies that $\Psi(\frac{1}{n},\frac{1}{n},\ldots,\frac{1}{n})\geq \Psi(1,0,\ldots,0)$.
The inequality $\Psi(\frac{1}{n},\frac{1}{n},\ldots,\frac{1}{n})\geq \Psi(\frac{1}{n-1},\ldots,\frac{1}{n-1}, 0)$,
which is the geometric analogue of the monotonicity of entropy in the Central Limit Theorem, was conjectured to hold in \cite{BMW11}.
However, it was shown in \cite{FMMZ16} (cf. \cite{FMMZ18}) that this inequality fails to hold, and therefore $\Psi$ cannot be Schur-concave,
for arbitrary Borel sets $B$. Note that if $B$ is convex, $\Psi$ is trivially Schur-concave, since it is a constant function equal to $\text{vol}_d(B)$.

Finally, it has recently been observed in \cite{WWM14:isit, MWW17:1, MWW17:2} that majorization ideas are very useful in understanding entropy power
inequalities in discrete settings, such as on the integers or on cyclic groups of prime order.

\section{Failure of Schur-concavity}\label{sec:schurfails}

Recall that a probability density $f$ on $\R$ is said to be {\it log-concave}
if it is of the form $f = e^{-V}$ for a convex function $V: \R \to \R\cup\{\infty\}$. Log-concave distributions 
emerge naturally from the interplay between information theory and convex geometry, and have recently
been a very fruitful and active topic of research (see the recent survey \cite{MMX17:0}). 

This section is devoted to a proof of Theorem \ref{thm:schurfails}, which in particular falsifies the Schur-concavity of $\Phi$
defined by \eqref{eq:Phi} even when the distribution under consideration is log-concave.

Let us denote 
\[
	Z_n = \frac{X_1+\ldots+X_n}{\sqrt{n}}.
\]
and let $p_n$ be the density of $Z_n$ and let $\varphi$ be the density of $Z$. Since $X_0$ is assumed to be log-concave, it satisfies $\E |X_0|^s < \infty$ for all $s>0$. According to the Edgeworth-type expansion described in \cite{BCG13:rate} (Theorem 3.2 in Chapter 3), we have (with any $m \leq s < m+1$)
\[
	(1+|x|^m)(p_n(x)-\varphi_m(x)) = o(n^{-\frac{s-2}{2}}) \qquad \textrm{uniformly in $x$},
\]
where 
\[
	\varphi_m(x) = \varphi(x) + \sum_{k=1}^{m-2} q_k(x) n^{-k/2}. 
\]	
Here the functions $q_k$ are given by
\[
	q_k(x) = \varphi(x) \sum H_{k+2j}(x) \frac{1}{r_1! \ldots r_k!} \left( \frac{\gamma_3}{3!} \right)^{r_1} \ldots  \left( \frac{\gamma_{k+2}}{(k+2)!} \right)^{r_k}, 
\]  
where $H_{n}$ are Hermite polynomials, 
\[
	H_n(x) = (-1)^n e^{x^2/2} \po{^n}{x^n} e^{-x^2/2},
\]
and the summation runs over all nonnegative integer solutions $(r_1,\ldots,r_k)$ to the equation $r_1 + 2r_2 + \ldots + k r_k = k$, and one uses the notation $j = r_1 +\ldots + r_k$. The numbers $\gamma_k$ are the cumulants of $X_0$, namely
\[
	\gamma_k = i^{-k} \po{^r}{t^r} \log \E e^{it X_0} \big|_{t=0}.
\]

Let us calculate $\varphi_4$. Under our assumption (symmetry of $X_0$ and $\E X_0^2=1$), we have $\gamma_3 = 0$ and $\gamma_4 = \E X_0^4 -3$. Therefore $q_1=0$ and 
\[
	q_2 = \frac{1}{4!}  \gamma_4  \varphi   H_4 = \frac{1}{4!} \gamma_4 \varphi^{(4)}, \qquad  \varphi_4 = \varphi +  \frac{1}{n} \cdot \frac{1}{4!}  (\E X_0^4 -3)  \varphi^{(4)}.
\]	
We get that  for any $\e \in (0,1)$
\[
	(1+x^4) (p_n(x) - \varphi_4(x)) = o(n^{-\frac{3-\e}{2}}), \qquad  \textrm{uniformly in $x$}.
\]

Let $f$ be the density of $X_0$.  Let us assume that it is of the form $f=\varphi+\delta$, where $\delta$ is even, smooth and compactly supported (say, supported in $[-2,-1] \cup [1,2]$) with bounded derivatives. Moreover, we assume that $\frac12 \varphi \leq f \leq 2\varphi$ and that $|\delta| \leq 1/4$. Multiplying $\delta$ by a very small constant we can ensure that $f$ is log-concave.  

We are going to use Theorem 1.3 from \cite{BCG16:2}. To check the assumptions of this theorem, we first observe that for any $\alpha > 1$ we have
\[
	D_\alpha(Z_1 || Z) = \frac{1}{\alpha-1} \log \left( \int \left(\frac{\varphi+\delta}{\varphi} \right)^{\alpha} \varphi \right) < \infty,
\] 
since $\delta$ has bounded support. We have to show that for sufficiently big $\alpha^\star = \frac{\alpha}{\alpha-1}$ there is 
\[
	\E e^{tX_0} < e^{\alpha^\star t^2 /2}, \qquad t \ne 0.  
\]
Since $X_0$ is symmetric, we can assume that $t>0$. Then 
\begin{align*}
	\E e^{tX_0} & = e^{t^2/2} + \sum_{k=1}^\infty \frac{t^{2k}}{(2k)!} \int x^{2k} \delta(x) \dd x \leq e^{t^2/2} + \sum_{k=1}^\infty \frac{t^{2k}}{(2k)!}  2^{2k} \int_{-2}^2 |\delta(x)| \dd x \\
	&< e^{t^2/2} + \sum_{k=1}^\infty \frac{(2t)^{2k}}{(2k)!}    = 1 +\sum_{k=1}^\infty \left( \frac{t^{2k}}{2^k k!} + \frac{(2t)^{2k}}{(2k)!} \right) \\
	&\leq  1 +\sum_{k=1}^\infty \left( \frac{t^{2k}}{k!} + \frac{(2t)^{2k}}{k!} \right) \leq \sum_{k=0}^\infty \frac{t^{2k} 4^{2k}}{k!}  = e^{16 t^2},  
\end{align*}
where we have used the fact that $\int \delta(x) \dd x = 0$, $\delta$ has a bounded support contained in $[-2,2]$ and $|\delta| \leq 1/4$. We conclude that
\[
	|p_n(x)-\varphi(x)| \leq \frac{C_0}{n} e^{-x^2/64}
\]
and thus
\[
	p_n(x) \leq \varphi(x) + \frac{C_0}{n} e^{-x^2/64} \leq C_0 e^{-x^2/C_0}.
\]
(In this proof $C$ and $c$ denote sufficiently large and sufficiently small universal constants that may change from one line to another. On the other hand, $C_0$, $c_0$ and $c_1$ denote constants that may depend on the distribution of $X_0$.)
Moreover, for $|x| \leq c_0 \sqrt{\log n}$ we have
\[
	p_n(x) \geq \varphi(x) - \frac{C_0}{n} e^{-x^2/64} \geq \frac{1}{n} e^{-x^2/64},
\]
so 
\[
	p_n(x) \geq  \frac{c_0}{n^{C_0}}, \qquad \textrm{for} \ |x| \leq c_0 \sqrt{\log n}.
\]
Let us define $h_n=p_n-\varphi_4$. Note that $\varphi_4=\varphi+\frac{c_1}{n} \varphi^{(4)}$, where $c_1=\frac{1}{4!}(\E X_0^4 -3)$. We have
\begin{align*}
	\int f \ast p_n \log f \ast p_n & = \int \left(f \ast \varphi + \frac{c_1}{n} f \ast \varphi^{(4)} + f \ast h_n \right) \log f \ast p_n  \\
	& = \int f \ast \varphi \log f \ast p_n  +  \frac{c_1}{n} \int  f \ast \varphi^{(4)} \log f \ast p_n   + \int f \ast h_n \log f \ast p_n  \\
	&= I_1 + I_2 + I_3.
\end{align*}
We first bound $I_3$. Note that 
\[
	(f \ast h_n)(x) \leq 2(\varphi \ast |h_n|)(x) \leq o(n^{-5/4} ) \int e^{-y^2/2} \frac{1}{1+(x-y)^4} \dd y.
\]	
Assuming without loss of generality that $x>0$, we have
\begin{align*}
	\int e^{-y^2/2} \frac{1}{1+(x-y)^4} \dd y & \leq  \int_{y \in [\frac12 x , 2x]} + \int_{y \notin [\frac12 x , 2x]} \\
	&\leq \int_{y \in [\frac12 x , 2x]}  e^{-x^2/8}  + \frac{1}{1+\frac{1}{16}x^4} \int_{y \notin [\frac12 x , 2x]} e^{-y^2/2}  \dd y \\
	& \leq \frac32 x e^{-x^2/8} + \frac{\sqrt{2\pi}}{1+\frac{1}{16}x^4} \leq \frac{C}{1+x^4}.  
\end{align*}
We also have 
\[
(f \ast p_n)(x) \leq 2(\varphi \ast p_n)(x) \leq C_0 (\varphi \ast e^{-y^2/C})(x) \leq C_0.
\]
Moreover, assuming without loss of generality that $x>0$,
\begin{align*}
	(f \ast p_n)(x) \geq \frac12 (\varphi \ast p_n)(x) &\geq \frac{c_0}{n^{C_0}} \int_{0 \leq y \leq c_0\sqrt{\log n}} e^{-(x-y)^2/2}  \\
	&\geq \frac{c_0}{n^{C_0}} \int_{0 \leq y \leq c_0\sqrt{\log n}} e^{-x^2/2} e^{-y^2/2} \geq \frac{c_0}{n^{C_0}} e^{-x^2/2}.
\end{align*}
Thus 
\[
	|\log f \ast p_n(x)| \leq \log(C_0 n^{C_0} e^{x^2/2}). 
\]
As a consequence
\begin{align*}
	I_3 \leq o(n^{-5/4}) \int  \frac{1}{1+x^4} \left|\log f \ast p_n(x)\right| \dd x &\leq  o(n^{-5/4}) \int  \frac{1}{1+x^4} \log(C_0 n^{C_0} e^{x^2/2}) \dd x \\
	&= o(n^{-5/4}\log n).  
\end{align*}
For $I_2$ fix $\beta>0$ and observe that
\begin{align*}
	\left| \int_{|x| \geq \beta \sqrt{\log n}} f \ast \varphi^{(4)} \log (f \ast p_n) \right| &\leq 2 \left| \int_{|x| \geq \beta \sqrt{\log n}} \varphi \ast |\varphi^{(4)}| \log (f \ast p_n) \right| \\
	&\leq C_0  \int_{|x| \geq \beta \sqrt{\log n}}  (1+x^4) e^{-x^2/4} \log(C_0 n^{C_0} e^{x^2/2}) \\
	&= o(n^{-c}).
\end{align*}
Hence,
\[
	\int  f \ast \varphi^{(4)} \log f \ast p_n = \int_{|x| \leq \beta \sqrt{\log n}} + \int_{|x| > \beta \sqrt{\log n}} = \int_{|x| \leq \beta \sqrt{\log n}} f \ast \varphi^{(4)} \log f \ast p_n + o(1).
\]
Writing $p_n = \varphi+r_n$ we get
\[
	\int_{|x| \leq \beta \sqrt{\log n}} f \ast \varphi^{(4)} \log f \ast p_n = \int_{|x| \leq \beta\sqrt{\log n}} (f \ast \varphi)^{(4)} \left[ \log( f \ast \varphi ) + \log \left( 1 + \frac{f \ast r_n}{f \ast \varphi}  \right) \right].
\]
Here 
\[
	\int_{|x| \leq \beta \sqrt{\log n}} (f \ast \varphi)^{(4)}  \log( f \ast \varphi ) = \int (f \ast \varphi)^{(4)}  \log( f \ast \varphi ) + o(1)
\]
and 
\[
	\int_{|x| \leq \beta \sqrt{\log n}} (f \ast \varphi)^{(4)}  \log \left( 1 + \frac{f \ast r_n}{f \ast \varphi}  \right)  = o(1),
\]
since for $|x| \leq \beta \sqrt{\log n}$ with $\beta$ sufficiently small we have
\[
	\left| \frac{f \ast r_n}{f \ast \varphi} \right| \leq  \frac{4}{n} \left| \frac{\varphi \ast (C_0e^{-x^2/{C_0}})}{\varphi \ast \varphi} \right| \leq \frac{C_0}{n} e^{C_0x^2} \leq \frac{C}{\sqrt{n}}. 
\]
By Jensen's inequality,
\[
	I_1  = \int f \ast \varphi \log f \ast p_n  \leq \int f \ast \varphi \log f \ast \varphi .
\]
Putting these things together we get
\[
	\int f \ast p_n \log f \ast p_n \leq \int f \ast \varphi \log f \ast \varphi + \frac{c_1}{n} \int (f \ast \varphi)^{(4)}  \log( f \ast \varphi ) + o(n^{-1}).
\]
This is
\[
	H(X_0+Z) \leq H(X_0+Z_n) + \frac{1}{n} \cdot \frac{1}{4!} (\E X_0^4 - 3) \int (f \ast \varphi)^{(4)}  \log( f \ast \varphi ) + o(n^{-1}).
\]
It is therefore enough to construct $X_0$ (satisfying all previous conditions) such that
\[
	(\E X_0^4 - 3) \int (f \ast \varphi)^{(4)}  \log( f \ast \varphi ) < 0.
\]

It actually suffices to construct $g$ such that $\int g = \int g x^2 = \int g x^4=0$  but the function $f = \varphi+\e g$ satisfies
\[
	\int (f \ast \varphi)'''' \log(f \ast \varphi) > 0 
\] 
for small $\e$. Then we perturb $g$ a bit to get $\E X_0^4 < 3$ instead of $\E X_0^4 = 3$. This can be done without affecting log-concavity.

 Let $\varphi_2 = \varphi \ast \varphi$. We have
\begin{align*}
	\int (f \ast \varphi)'''' \log (f \ast \varphi) & = \int (\varphi_2 + \e \varphi \ast g )'''' \log (\varphi_2 + \e \varphi \ast g ) \\
	& \int (\varphi_2 + \e \varphi \ast g )'''' \left( \log (\varphi_2) + \e  \frac{\varphi \ast  g}{\varphi_2} - \frac12 \e^2 \left(\frac{\varphi \ast  g}{\varphi_2} \right)^2    \right). 
\end{align*}  
The leading term  and the term in front of $\e$ vanish (thanks to $g$  being orthogonal to $1,x,\ldots,x^4$). The term in front of $\e^2$ is equal to
\[
	J = \int \frac{(\varphi \ast g)'''' (\varphi \ast g)}{\varphi_2} - \frac12 \int \frac{\varphi_2''''(\varphi \ast g)^2}{\varphi_2^2} = J_1 - J_2. 
\]
The first integral is equal to
\[
	J_1 = \int \int \int 2\sqrt{\pi} e^{x^2/4} g''''(s)g(t) \frac{1}{2\pi} e^{-(x-s)^2/2} e^{-(x-t)^2/2}   \dd x \dd s \dd t. 
\]
Now,
\[
	\int  2\sqrt{\pi} e^{x^2/4} \frac{1}{2\pi} e^{-(x-s)^2/2} e^{-(x-t)^2/2}   \dd x  =  \frac{2 e^{\frac{1}{6} \left(-s^2+4 s t-t^2\right)}}{\sqrt{3}}.
\]
Therefore,
\[
	 J_1 = \frac{2}{\sqrt{3}}  \int \int  e^{\frac{1}{6} \left(-s^2+4 s t-t^2\right)} g''''(s) g(t)  \dd s \dd t .
\]
If we integrate the first integral $4$ times by parts we get
\begin{align*}
	J_1 = \frac{2}{81 \sqrt{3}} \int \int e^{\frac16 (-s^2 + 4 s t - t^2)} &\Big[27 + s^4 - 8 s^3 t - 72 t^2 \\
	&+ 
   16 t^4 - 8 s t (-9 + 4 t^2) + 6 s^2 (-3 + 4 t^2)\Big] g(s) g(t) \dd s \dd t
\end{align*}
Moreover,
\[
	\frac{\varphi_2''''}{\varphi_2^2} = \frac{\sqrt{\pi}}{16} (12 - 12 x^2 + x^4) e^{x^2/4},
\]
so we get
\[
	J_2 =  \int \int \int  \frac{\sqrt{\pi}}{16} (12 - 12 x^2 + x^4) e^{x^2/4} g(s) g(t) \frac{1}{2\pi} e^{-(x-s)^2/2} e^{-(x-t)^2/2}  \dd x \dd s \dd t.
\]
Since
\begin{align*}
	\int \frac{\sqrt{\pi}}{16} (12 - 12 x^2 + x^4) e^{x^2/4} & \frac{1}{2\pi} e^{-(x-s)^2/2} e^{-(x-t)^2/2}  \dd x \\
	 &=   \frac{1}{81 \sqrt{3}} e^{\frac16 (-s^2 + 4 s t - 
    t^2)} \left[27 + (s + t)^2 (-18 + (s + t)^2) \right] ,
\end{align*}
we arrive at
\[
	J_2 = \int \int \frac{1}{81 \sqrt{3}} e^{\frac16 (-s^2 + 4 s t - 
    t^2)} \left[27 + (s + t)^2 (-18 + (s + t)^2) \right] g(s) g(t) \dd s \dd t.
\]
Thus $J = J_1 - J_2$ becomes
\begin{align*}
	J =\frac{1}{81 \sqrt{3}}  \int  \int e^{\frac16 (-s^2 + 4 s t - t^2)} & \Big[27 + s^4 - 20 s^3 t - 126 t^2 + 31 t^4 \\
	&+ 6 s^2 (-3 + 7 t^2) + 
 s (180 t - 68 t^3) \Big] g(s) g(t) \dd s \dd t.
\end{align*}
The function
\[
	g(s) = \left(\frac{7280}{69} |s|^3 -\frac{11025}{23} s^2 + \frac{49000}{69} |s| -\frac{7875}{23}\right)\1_{[1,2]}(|s|) 
\]
satisfies $\int g = \int g x^2 = \int g x^4 = 0$. 
Numerical computations show that for this $g$, $J > 0.003$.
\hfill$\square$

\section{Entropy power inequalities under symmetries}\label{sec:proofcomplex}

The heart of the folklore proof of $h(S_2)\geq h(S_1)$ for symmetric distributions (see, e.g., \cite{WM14})
is that for possibly dependent random variables $X_1$ and $X_2$, the $SL(n,\R)$-invariance of differential entropy
combined with subadditivity imply that
\begin{equation*}\begin{split}
h(X_1, X_2)&= h\bigg(\frac{X_1+X_2}{\sqrt{2}}, \frac{X_1-X_2}{\sqrt{2}}\bigg)\\
&\leq   h\bigg(\frac{X_1+X_2}{\sqrt{2}}\bigg)+h\bigg(\frac{X_1-X_2}{\sqrt{2}}\bigg).
\end{split}\end{equation*}
If the distribution of $(X_1, X_2)$ is the same as that of $(X_1, -X_2)$, we deduce that
\begin{equation}\label{eq:hj2}
h\bigg(\frac{X_1+X_2}{\sqrt{2}}\bigg)\geq \frac{h(X_1, X_2)}{2} .
\end{equation}
If, furthermore, $X_1$ and $X_2$ are i.i.d., then $h(X_1, X_2)=2h(X_1)$, yielding $h(S_2)\geq h(S_1)$.
Note that under the i.i.d. assumption, the requirement that the distributions of $(X_1, X_2)$ and $(X_1, -X_2)$ coincide
is equivalent to the requirement that $X_1$ (or $X_2$) has a symmetric distribution.

Without assuming symmetry but assuming independence, we can use the fact from \cite{KM14} 
that $h(X-Y)\leq 3h(X+Y)-h(X)-h(Y)$ for independent random variables $X, Y$ to deduce
$\frac{1}{2} [ h(X_1)+h(X_2)] \leq h\big(\frac{X_1+X_2}{\sqrt{2}}\big) +\frac{1}{4}\log 2$.
In the i.i.d. case, the improved bound $h(X-Y)\leq 2h(X+Y)-h(X)$ holds \cite{MK10:isit}, which implies $h(X_1) \leq h\big(\frac{X_1+X_2}{\sqrt{2}}\big) +\frac{1}{6}\log 2$.
These bounds are, however, not particularly interesting since they are weaker than the classical entropy power inequality;
if they had recovered it, these ideas would have represented by far its most elementary proof.

Hao and Jog \cite{HJ18} generalized the inequality \eqref{eq:hj2} to the case where one
has $n$ random variables, under a natural $n$-variable extension of the distributional requirement,
namely unconditionality. However, they used a proof that goes through Fisher information inequalities,
similar to the original Stam proof of the full entropy power inequality. 
The main observation of this section is simply that under certain circumstances, one can give
a direct and simple proof of the Hao--Jog inequality, as well as others like it, akin to the 2-line proof
of the inequality \eqref{eq:hj2} given above. The ``certain circumstances'' have to do with the existence
of appropriate linear transformations that respect certain symmetries-- specifically Hadamard matrices.

Let us first outline how this works in the real case. Suppose $n$ is a dimension for which there exists a {\it Hadamard matrix}--
namely, a $n\times n$ matrix with all its entries being 1 or $-1$, and its rows forming an orthogonal set of vectors.
Dividing each row by its length $\sqrt{n}$ results in an orthogonal matrix $O$, all of whose entries are $\pm\frac{1}{\sqrt{n}}$.
By unconditionality, each coordinate of the vector $OX$ has the same distribution as $\frac{X_1+\ldots+X_n}{\sqrt{n}}$. 
Hence
\[
\Ent{X} = \Ent{OX} \leq \sum_{j=1}^n \Ent{(OX)_j} = n\Ent{\frac{X_1+\ldots+X_n}{\sqrt{n}}},
\]
where the inequality follows from subadditivity of entropy. This is exactly the Hao-Jog inequality for those dimensions where
a Hadamard matrix exists. It would be interesting to find a way around the dimensional restriction, but we do not
currently have a way of doing so.

As is well known, other than the dimensions $1$ and $2$, Hadamard matrices may only exist for dimensions that are multiples of 4.
As of this date, Hadamard matrices are known to exist for all multiples of 4 up to 664 \cite{KT05:2}, and it is a major open problem whether 
they in fact exist for all multiples of 4. (Incidentally, we note that the question of existence of Hadamard matrices can actually be formulated in the entropy language. 
Indeed, Hadamard matrices are precisely those that saturate the obvious bound for the entropy of an orthogonal matrix \cite{GMPS03}.)

In contrast, complex Hadamard matrices exist in every dimension. A {\it complex Hadamard matrix} of order $n$ is a $n\times n$ matrix with
complex entries all of which have modulus 1, and whose rows form an orthogonal set of vectors in $\mathbb{C}^n$.  
To see that complex Hadamard matrices always exist, we merely exhibit the Fourier matrices, which are a well known example of them:
these are defined by the entries
$
H_{j,k}=\exp\{ \frac{2\pi i (j-1)(k-1)}{n}\} ,
$
for $j, k=1, \ldots, n$, and are related to the  discrete Fourier transform (DFT) matrices.
Complex Hadamard matrices play an important role in quantum information theory \cite{TZ06}. 
They also yield Theorem \ref{thm:complex}.

\vspace{.1in}
\noindent{\it Proof of Theorem \ref{thm:complex}.}
Take any $n \times n$ unitary matrix $U$ which all entries are complex numbers of the same modulus $\frac{1}{\sqrt{n}}$;
such matrices are easily constructed by multiplying a complex Hadamard matrix by $n^{-1/2}$.
(For instance, one could take $U = \frac{1}{\sqrt{n}}[e^{2\pi ikl/n}]_{k,l}$.)
By complex-unconditionality, each coordinate of the vector $UX$ has the same distribution, the same as $\frac{X_1+\ldots+X_n}{\sqrt{n}}$. Therefore, by subadditivity,
\[
\Ent{X} = \Ent{UX} \leq \sum_{j=1}^n \Ent{(UX)_j} = n\Ent{\frac{X_1+\ldots+X_n}{\sqrt{n}}},
\]
which finishes the proof. 
\hfill$\square$
\vspace{.1in}

Let us mention that the invariance idea above also very simply yields the inequality
$$
D(X) \leq \frac{1}{2} |h(X_1+X_2)-h(X_1-X_2)| ,
$$
where $D(X)$ denotes the relative entropy of the distribution of $X$ from the closest Gaussian (which is the one
with matching mean and covariance matrix), and $X_1, X_2$ are independent copies of a random vector $X$ in $\R^n$.
First observed in \cite[Theorem 10]{MK18}, this fact quantifies the distance from Gaussianity of a random vector in terms of
how different the entropies of the sum and difference of i.i.d. copies of it are.

Finally, we mention that the idea of considering two i.i.d. copies and using invariance (sometimes
called the ``doubling trick'') has been used in sophisticated ways as a key tool to study both functional inequalities \cite{Lie90, Car91, Bar98a}
and problems in network information theory (see, e.g., \cite{GN14, Cou18:1}).

\section*{Acknowledgments}

We learned about the Edgeworth expansion used in our proof of Theorem \ref{thm:schurfails} from S. Bobkov during the AIM workshop \emph{Entropy power inequalities}. 
We are immensely grateful to him as well as the AIM and the organisers of the workshop which was a valuable and unique research experience.


\end{document}